\documentclass[showpacs,aps,superscriptaddress,prb,twocolumn,amsmath]
{revtex4}
\usepackage{txfonts}
\usepackage{color}
\usepackage{amssymb}
\usepackage{graphicx}% Include figure files
\usepackage{dcolumn}% Align table columns on decimal point
\usepackage{bm}% bold math
\usepackage{hyperref}
\begin{document}
\preprint{APS/123-QED}
\title{Simple Cubic Carbon Phase C21-sc: A Promising Superhard Carbon Conductor}
%====================================================================
\author{Chaoyu He}
\affiliation{Hunan Key Laboratory for Micro-Nano Energy Materials
and Devices, Xiangtan University, Hunan 411105, P. R. China;}
\affiliation{School of Physics and Optoelectronics, Xiangtan
University, Xiangtan 411105, China.}
\author{Lijun Meng}
\affiliation{Hunan Key Laboratory for Micro-Nano Energy Materials
and Devices, Xiangtan University, Hunan 411105, P. R. China;}
\affiliation{School of Physics and Optoelectronics, Xiangtan
University, Xiangtan 411105, China.}
\author{Chao Tang}
\affiliation{Hunan Key Laboratory for Micro-Nano Energy Materials
and Devices, Xiangtan University, Hunan 411105, P. R. China;}
\affiliation{School of Physics and Optoelectronics, Xiangtan
University, Xiangtan 411105, China.}
\author{Jianxin Zhong}
\email{jxzhong@xtu.edu.cn}\affiliation{Hunan Key Laboratory for
Micro-Nano Energy Materials and Devices, Xiangtan University, Hunan
411105, P. R. China;} \affiliation{School of Physics and
Optoelectronics, Xiangtan University, Xiangtan 411105, China.}
\date{\today}
%===============================================================
\begin{abstract}
Traditionally, all superhard carbon phases including diamond are
electric insulators and all conductive carbon phases including
graphite are mechanically soft. Based on first-principles
calculation results, we report a superhard but conductive carbon
phase C21-sc which can be obtained through increasing the sp$^3$
bonds in the previously proposed soft and conductive phase C20-sc
(\textbf{Phys. Rev. B 74, 172101 2006}). We also show that further
increase of sp$^3$ bonds in C21-sc results in a superhard and
insulating phase C22-sc with sp$^3$ bonds only. With C20-sc, C21-sc,
C22-sc and graphite, the X-ray diffraction peaks from the
unidentified carbon material synthesized by compressing the mixture
of tetracyanoethylene and carbon black (\textbf{Carbon, 41, 1309,
2003}) can be understood. In view of its positive stability,
superhard and conductive features, and the strong possibility of
existence in previous experiments, C21-sc is a promising
multi-functional material with potential applications in extreme conditions.\\
\end{abstract}
\maketitle
%section{Introduction}
\indent Superhard materials with extreme Vicker's hardness larger
than 40 GPa are widely used in mechanical industry. For instance,
diamond and cubic boron nitride with extreme hardness are often used
as tools for cutting and polishing other materials. However, both
diamond and cubic boron nitride are electronic insulating. Most of
other superhard materials are also electronic insulating. In the
past decade, many efforts have been paid on searching for superhard
and conductive materials \cite{sm1,sm2,sm3,sm4,sm5} for
multi-functional applications in extreme conditions. For example,
the boron-doped diamond \cite{sm1,sm5} can be superhard and
conductive due to its diamond structure for superhard mechanical
property and the addition of the necessary electrons for
conductivity from boron.\\
%*************************figure 1****,bb=0 0 1750 400*********************************************************************
\begin{figure}
\center
\includegraphics[width=3.5in]{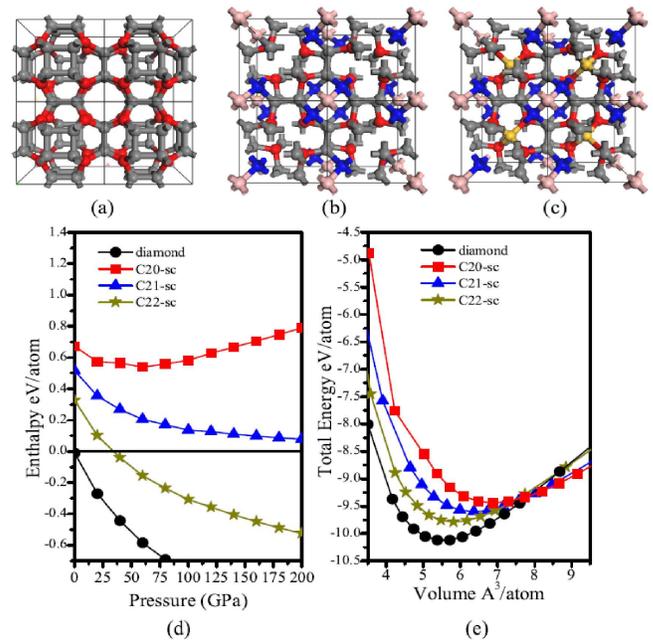}\\
\caption{Perspective view of C20-sc (a), C21-sc (b) and C22-sc (c)
in their crystalline cells. Balls in different colors indicate
different atoms (Their positions 1, 2, 3, 4 and 5 listed in Table I
are correspondingly marked in red, blue, grey, brown and yellow,
respectively.); E-P cures (d) of diamond, C20-sc, C21-sc and C22-sc
relative to graphite and E-V cures (e) of diamond, C20-sc, C21-sc
and C22-sc.}\label{fig1}
\end{figure}
\indent Carbon with broad sp, sp$^2$ and sp$^3$ hybridizing ability
can form numerous carbon allotropes, including the superhard diamond
and the very soft graphite. Superhard carbon phases (diamond and its
allotropes) are optically transparent insulators and conductive
carbon phases (graphite and its allotropes) are mechanically soft.
The superhardness of diamond and its allotropes is mainly
contributed by the entire covalent sp$^3$ short bonds, which result
in insulating electronic properties. Based on numerous previous
investigations on the mechanical and electronic properties of
different carbon allotropes, one may conclude that carbon allotropes
with entire sp$^3$ hybridization \cite{M,W,Z3,PCCP,HS,JSM,zq} are
always mechanically superhard and electronically insulating. Carbon
allotropes with entire sp$^2$ hybridization
\cite{bct-4,H6,H62,C20,cg,sp21} are always mechanically soft, but
they are not always metallic. For example, one has soft and metallic
bct-4 \cite{bct-4}, C20 \cite{C20} and H6-carbon \cite{H6,H62}, as
well as soft and insulating polybenzene \cite{cg} and sp$^2$-diamond
\cite{sp21}. All the previously proposed carbon phases with mixed
sp$^2$ and sp$^3$ bonds \cite{sp2sp3,g1,hm,jx, c20-sc} are soft
materials which can be metals or insulators. There is no carbon
phase with both superhard and conductive features in the published
literature. In this letter, we report a simple cubic carbon phase
C21-sc which is superhard and conductive. C21-sc can be constructed
by increasing the sp$^3$ bonds in the previously proposed metallic
carbon phase C20-sc \cite{c20-sc} with both sp$^2$ and sp$^3$ bonds.
Further increase of sp$^3$ bonds in C21-sc results in a superhard
but insulating phase C22-sc with sp$^3$ bonds only. We show that
with C20-sc, C21-sc, C22-sc and graphite, the X-ray diffraction
peaks found in an
unidentified carbon material \cite{c10} can be understood.\\
%=============================================================================================
%*************************figure 2*************************************************************************
\begin{figure}
\includegraphics[width=3.5in]{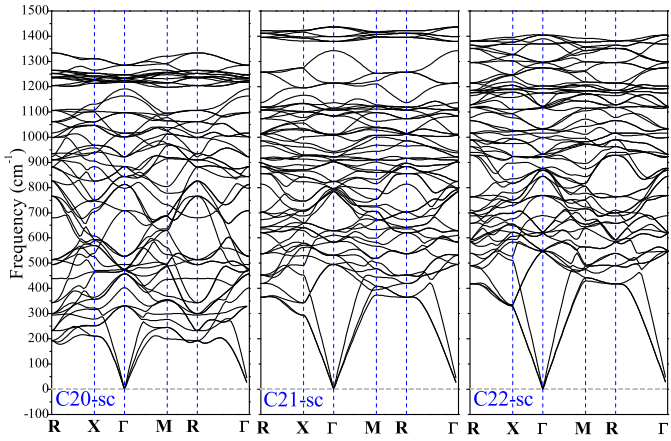}\\
\caption{Phonon band structures of C20-sc, C21-sc and
C22-sc.}\label{fig2}
\end{figure}
\indent All calculations of structural optimization and properties
are carried out using the density functional theory within local
density approximation (LDA) \cite{13, 14} as implemented in Vienna
ab initio simulation package (VASP) \cite{16, 17}. The interactions
between nucleus and the 2s$^{2}$2p$^{2}$ valence electrons of carbon
are described by the projector augmented wave (PAW) method \cite{18,
19}. A plane-wave basis with a cutoff energy of 500 eV is used to
expand the wave functions and the Brillouin Zone (BZ) sample meshes
are set to be dense enough (less than 0.21 {\AA}$^{-1}$) to ensure
the accuracy of calculations. The structures of C20-sc, C21-sc,
C22-sc and diamond are fully optimized up to the residual force on
every atom less than 0.005 eV/{\AA}. The vibrational properties of
C20-sc, C21-sc and C22-sc are investigated by using the PHONON
package \cite{21} with the forces calculated from VASP to evaluate
their dynamical stability. The three inequivalent elastic constants
(C$_{11}$, C$_{12}$ and C$_{44}$) of diamond, C20-sc, C21-sc and
C22-sc are calculated as the second-order coefficients in the
polynomial function of distortion parameter $\delta$ used to fit
their total energies according to the Hooke's law. Three groups of
deformations, namely, (e$_{1,2}$=$\delta$,
e$_3$=(1+$\delta$)$^{-2}$-1, e$_{4,5,6}$=0), (e$_{1,2,3}$=$\delta$,
e$_{4,5,6}$=0), and (e$_6$=$\delta$,
e$_3$=$\delta$$^2$(4-$\delta$$^2$)$^{-1}$, e$_{1,2,5}$=0) are
considered in each cubic phase. The bulk modulus (\textbf{B}) and
shear modulus (\textbf{G}) are evaluated according to Hill's formula
\cite{hill} based on the calculated elastic constants. To further
analyze the hardness of these carbon allotropes, we adopt the
recently introduced empirical scheme \cite{hard} to evaluate
Vicker's hardness (H${_v}$) determined by
\textbf{B} and \textbf{G} as \textbf{H${_v}$=2(G${^3}$/B${^2}$)${^{0.585}}$-3}.\\
%++++++++++++++++++++
\indent C20-sc contains 20 carbon atoms in its cubic cell (Pm-3m)
with lattice constant of 5.163 {\AA}. It has two inequivalent carbon
atoms, namely, sp$^2$ and sp$^3$ carbon atoms locating at positions
of 8g (0.238, 0.238, 0.238) and 12i (0.00, 0.651, 0.651),
respectively. The eight sp$^2$ carbon atoms symmetrically distribute
on the body diagonal of the cubic cell, which role as inter-linkers
between the surface squares formed by the sp$^3$ carbon atoms. There
are two inequivalent carbon bonds in C20-sc, namely, the
inner-square ones B$_{33}$ containing only sp$^3$ carbon atoms (we
mark bonds as B$_{ij}$ and angles as $\theta$$_{ijk}$, where i, j
and k mean different types of carbon atoms as indicated in Table I.)
and the inter-square ones B$_{13}$ linking the sp$^2$ and sp$^3$
carbon atoms. The corresponding bond lengthes are B$_{33}$=1.564
{\AA} and B$_{13}$=1.471 {\AA}, which are close to those in diamond
and graphite, respectively. The four distinct bond angles in C20-sc
are $\theta$$_{333}$=90$^o$, $\theta$$_{331}$=112.95$^o$,
$\theta$$_{131}$=113.06$^o$ and $\theta$$_{313}$=120$^o$, respectively.\\
%+++++++++++++table I ++++++++++++++++++++++++++++++++%
\begin{table*}
  \centering
  \caption{Fundamental structural information, cohesive energies, elastic constants and mechanical properties of diamond, C20-sc, C21-sc and C22-sc.}\label{tabI}
\begin{tabular}{c c c c c c}
\hline \hline
Items       &Diamond         &C20-sc         &C21-sc         &C22-sc\\
\hline
Space group &Fd-3m (No.227)  &Pm-3m (No.221)  &P-43m (No.215)  &P-43m (No.215)\\
Lattice constant&3.536 {\AA} &5.163 {\AA}     &5.112{\AA}      &5.097{\AA}\\
Mass density &3.611 Mg/cm$^3$&2.899 Mg/cm$^3$ &3.134 Mg/cm$^3$ &3.313 Mg/cm$^3$\\
Cohesive energy &-8.989       & -8.315        &-8.471          &-8.661           \\
Position 1 (red)  &8a:0.000,0.000,0.000&8g:0.238,0.238,0.238 &4e:0.278,0.278,0.278&4e:0.304,0.304,0.304\\
Position 2 (blue)    &-                &-                    &4e:0.176,0.826,0.176&4e:0.176,0.826,0.176\\
Position 3 (grey)    &-                &12i:0.00, 0.651,0.349&12i:0.000,0.651,0.349&12i:0.981,0.649,0.351\\
Position 4 (brown)   &-                &-                    &1a:0.000,0.000,0.000&1a:0.000,0.000,0.000\\
Position 5 (yellow)  &-                &-                    &-                    &1b:0.500,0.500,0.500\\
C$_{11}$       &1100.46 GPa      &577.13 GPa      &754.73 GPa        &959.79 GPa\\
C$_{12}$       &149.73 GPa       &275.89 GPa      &168.61 GPa        &99.46 GPa\\
C$_{44}$       &589.91 GPa       &243.89 GPa      &370.99 GPa        &453.97 GPa\\
Shear modulus  &541.07 GPa       &195.55 GPa      &337.57 GPa        &444.29 GPa\\
Bulk modulus   &466.64 GPa       &369.64 GPa      &363.98 GPa        &386.24 GPa\\
Verker's hardness&91.44 GPa      &17.79 GPa       &52.19 GPa         &80.38 GPa  \\
\hline \hline
\end{tabular}
\end{table*}
\indent With standard sp$^2$ hybridization and distorted sp$^3$
hybridization, C20-sc possesses relatively high energy. Its cohesive
energy of -8.315 eV/atom is 674 meV higher than that of diamond.
However, C20-sc is more favorable than the experimentally viable
graphdiyne. Its cohesive energy is about 300 meV per atom lower than
that of graphdiyne (-8.083 eV/atom), indicating that it is
experimentally viable too. Moreover, our calculations on its
vibrational property show that there is no imaginary frequency in
its phonon band structure (see in Fig.2), which confirms that C20-sc
is a dynamically stable. Its calculated elastic constants (see Tab
I) also satisfy the mechanical stability criteria of cubic lattice,
suggesting that C20-sc is mechanically stable too. The phase
stability of C20-sc can also be confirmed by the quadratic E-V
relation near the equilibrium V$_0$ as shown in Fig.1 (e).\\
\indent As a cage-like structure with the ratio of 2:3 for sp$^2$
and sp$^3$ bonds, C20-sc is a sparse, soft and conductive material.
From Fig.1 (e), we can see that C20-sc possesses a relatively larger
equilibrium volume (V$_0$). Its mass density of 2.899 Mg/cm$^3$ is
just a little higher than that of graphite and its verker's hardness
is very low (17.79 GPa) in comparison with that of diamond. From the
calculated electronic band structures (EBS) and projected energy
density of states (PDOS) of C20-sc (see Fig.3), we can see that
C20-sc is a metallic phase with states around and crossing the
Fermi-level. The PDOS shows that the metallic states around the
Fermi-level in C20-sc are mainly contributed from the sp$^2$
hybridized carbon atoms. While the sp$^3$ hybridized carbon atoms
contribute states locating at relatively lower energy area. From the
bonding charge density of C20-sc shown in Fig.3, we can see that
bonding charge in C20-sc distribute around sp$^3$ and
sp$^2$ carbon atoms with tetrahedral and triangular patterns, respectively.\\
\indent Based on the relation between the structure and electronic
property of C20-sc as well as the common knowledge that more sp$^3$
hybridization leads to higher hardness, we modify the structure of
C20-sc to construct the new carbon phase C21-sc which can be
superhard and conductive. As shown in Fig.1, we construct C21-sc
through adding carbon atoms (colored in brown) in C20-sc at the
vertex positions of 1a (0.000, 0.000, 0.000) and moving 4 of the 8
sp$^2$ carbon atoms (colored in blue), from 4e (0.238, 0.238, 0.238)
to 4e (0.176, 0.176, 0.176) along the body diagonal toward to the
vertex to enclose the added vertex atoms in standard tetrahedrons
with proper size, for standard sp$^3$ hybridization. Such an
operation reduces the symmetry of the system from Pm-3m to F-43m and
changes C20-sc to C21-sc with reduced sp$^2$ carbon atoms (with the
ratio of 4:17 for sp$^2$:sp$^3$). After optimization, the lattice
constant of C21-sc is 5.112 {\AA}. From table I, we can see that
with the added atoms at 1a position, the atoms at 8g position in
C20-sc decompose into two inequivalent groups locating at two
inequivalent positions of 4e (0.278, 0.278, 0.278) and 4e (0.176,
0.176, 0.176) with sp$^2$ and sp$^3$ hybridizations, respectively.
It should be noticed that carbon atoms at the sp$^2$ positions
slightly move toward the vertex (from (0.238, 0.238, 0.238) in
C20-sc to (0.278, 0.278, 0.278) in C21-sc) after optimization,
forming a non-planar sp$^2$ configuration. The sp$^3$ carbon atoms
still locate at the 12i
position of (0.000, 0.651, 0.349). \\
\indent There are four distinct C-C bonds in C21-sc which are
B$_{33}$=1.540 {\AA}, B$_{13}$=1.431 {\AA}, B$_{23}$=1.602 {\AA} and
B$_{24}$=1.562 {\AA}. Three equivalent bond angles around atom 1 are
formed by the triangularly distributed atoms 3. They are sp$^2$-like
bond angles of 114.5$^o$ ($\theta$$_{313}$), which is slightly
smaller than that of the standard sp$^2$ hybridization (120$^o$).
Two inequivalent bond angles around atom 2 are $\theta$$_{323}$ and
$\theta$$_{324}$ of 112.29$^o$ and 106.48$^o$, respectively. The
four inequivalent bond angles around atom 3 are
$\theta$$_{333}$=89.26$^o$, $\theta$$_{133}$=111.46$^o$,
$\theta$$_{233}$=118.62$^o$ and $\theta$$_{132}$=106.79$^o$. Four
equivalent atoms 2 tetrahedrally distribute around atom 4, forming
six equivalent sp$^3$ hybridized bond angles $\theta$$_{242}$ of 109.47$^o$.\\
%*************************figure 2*************************************************************************
\begin{figure}
\includegraphics[width=3.5in]{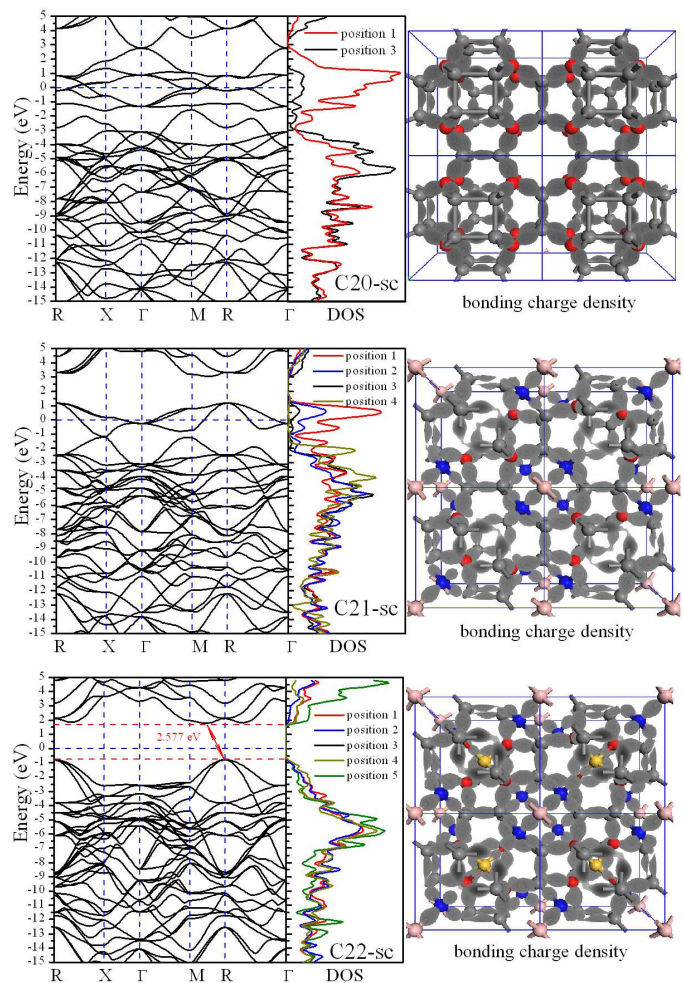}\\
\caption{Electronic band structures (left), projected energy density
of states (middle) and bonding charge density (right)  of C20-sc
(top), C21-sc (center) and C22-sc (bottom).}\label{fig3}
\end{figure}
\indent With the increased sp$^3$ hybridization, C21-sc is
energetically more favorable than C20-sc. Its cohesive energy of
-8.471 eV/atom is about 160 meV lower than that of C20-sc. As shown
in Fig.1 (d), its energetic stability gradually approaches to that
of graphite as the external pressure increases. Fig.2 shows the
phonon band structure of C21-sc, which confirms the positive
dynamical stability of C21-sc. The three independent elastic
constants of C21-sc with C$_{11}$=754.73 GPa, C$_{12}$=168.61 GPa
and C$_{44}$=370.99 GPa satisfy the mechanical stability criteria,
indicating that C21-sc is mechanically stable too. In Fig.1 (e), the
phase stability of C21-sc
is also confirmed by the quadratic E-V relation near the equilibrium V$_0$.\\
\indent C21-sc still contains 4 sp$^2$ hybridized carbon atoms per
cubic cell. Its mass density of 3.134 Mg/cm$^3$ is just a little
higher than that of C20-sc. As expected, the increased sp$^3$ ratio
makes C21-sc harder than C20-sc. In fact, C21-sc is a superhard
material in view of its high verker's hardness of 52.19 GPa. It is
exciting that C21-sc is electrically conductive. From the calculated
EBS and PDOS of C21-sc, we can see that the reduction of sp$^2$
hybridized carbon atoms reduces the numbers of the energy bands
around Fermi-level. However, C21-sc still behaves as a classical
metal with obvious electronic bands crossing the Fermi-level. From
the PDOS shown in Fig.3, the metallic states around the Fermi-level
are mainly contributed by the sp$^2$ hybridized atoms 1. We can also
see that the bonding charge density (Fig.3) distributes on the
sp$^3$ atoms with tetrahedral configurations and on the sp$^2$
atoms with triangular configurations.\\
%***********************************
\indent Further modification of C20-sc by adding carbon atoms
(colored in yellow) at the body center 1b (0.500, 0.500, 0.500) in
C21-sc provides us a new superhard insulating phase C22-sc. Detailed
structural information of C22-sc is given in Tab I. After
optimization, the lattice constant of C22-sc is reduced, in
comparison with that of C21-sc, to 5.097 {\AA}. With the entirely
 sp$^3$ hybridization, C22-sc is energetically more favorable than C20-sc and
C21-sc. Its cohesive energy of -8.661 eV/atom is about 180 meV lower
than that of C21-sc. Specially, C22-sc becomes more stable than
graphite when the external pressure increases to 34 GPa, which is in
good agreement with the fact that high pressure prefers saturated
bonds than unsaturated ones \cite{exp5,jacs1,jpcm}. Both the
dynamical and mechanical stabilities of C22-sc are confirmed to be
positive by its
phonon band structure and elastic constants, respectively.\\
\indent With sp$^3$ carbon atoms only, C22-sc becomes denser and
harder than C20-sc and C21-sc. Its mass density of 3.313 Mg/cm$^3$
is slightly smaller than that of diamond and its verker's hardness
of 80.38 GPa is comparable to that of diamond. The whole sp$^3$
hybridization makes C22-sc an insulator. As shown in Fig.3, it
possesses an indirect band gap of 2.577 eV. The bonding charge
density in C22-sc tetrahedrally distributes on the sp$^3$
bonds around every carbon atom.\\
%++++++++++++++++++++
\begin{figure}
\includegraphics[width=3.5in]{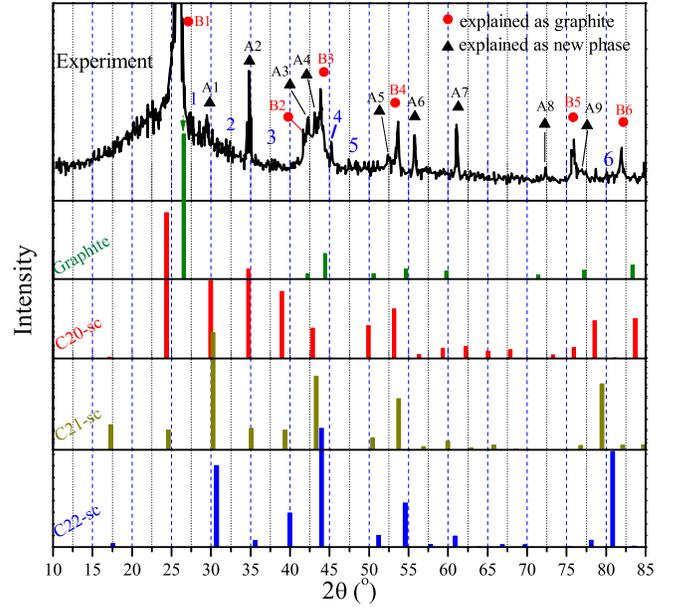}\\
\caption{Experimental X-ray diffraction patterns of the new carbon
material \cite{c10}(black lines) and simulated X-ray diffraction
peaks of graphite (green), C20-sc (red), C21-sc (dark yellow) and
C22-sc (blue).}\label{fig4}
\end{figure}
%%_________________________________
\indent The above theoretical results show that C21-sc possesses
both superhard and conductive properties, which can be potentially
used as a multi-functional material in extreme conditions. In fact,
in a previous experimental work in 2003 \cite{c10}, a pure carbon
material was synthesized by compressing the mixture of
tetracyanoethylene and carbon black. The new carbon material
possesses a simple cubic lattice with lattice constant of 5.14
{\AA}, which is very close to those of C20-sc, C21-sc and C22-sc. In
the experimental work, it is believed that six X-ray diffraction
peaks (B1-B6 marked as red solid circles) are contributed from
graphite, while nine new peaks (marked as A1-A9 with black solid
triangles in Fig.4) are contributed by an unknown new carbon phase.
After simulation of the X-ray diffraction of C20-sc, C21-sc and
C22-sc, we find that the experimental X-ray diffraction peaks can be
understood by the contribution of graphite, C20-sc, C21-sc and
C22-sc. As shown in Fig.4, we believe that peaks A1, A2 and A4 are
from C20-sc, A3 from graphite, A7 from C22-sc and A9 from C21-sc.
From our understanding, peaks B1-B6 are not from graphite. Although
the position of B1 corresponds to graphite, its width can not be
understood solely with graphite. We find that the existence of
C20-sc and C21-sc can well explain the width of peak B1 and it is
better to explain B3 with C22-sc, B4 and B5 with C20-sc and B6 with
C21-sc rather than graphite. Especially, we find that all the X-ray
diffraction peaks of graphite, C20-sc, C21-sc and C22-sc can be used
to explain other peaks found in the experiment. The excellent match
of lattice constants and X-ray diffraction peaks suggest the
synthesized material contain graphite, C20-sc, C21-sc and C22-sc.\\
%++++++++++++++++++++++++++++++++++++++++++++++++++++++++++++++++++++++++++++++
\indent In summary, through modification of the soft and conductive
carbon phase C20-sc, we have predicted two new cubic carbon phases
C21-sc and C22-sc which possess viable energetic stability, positive
dynamical and mechanical stabilities. Both C21-sc and C22-sc are
superhard materials similar to diamond. However, C21-sc is
electrically conductive while C22-sc is insulating. We show that the
X-ray diffraction peaks in an unknown cubic carbon material
synthesized by compressing a mixture of tetracyanoethylene and
carbon black can be understood from the characteristic diffraction
peaks of C20-sc, C21-sc, C22-sc and graphite. The strong theoretical
and experimental evidences of the existence of superhard and
conductive carbon C21-sc may open a new window for the study of
multi-functional
materials in extreme conditions.\\
%\section*{Acknowledgement}
\indent This work is supported by the National Natural Science
Foundation of China (Grant Nos. A040204 and 11204261), the National
Basic Research Program of China (2012CB921303 and 2015CB921103), the
Hunan Provincial Innovation Foundation for Postgraduate (Grant No.
CX2013A010), the Young Scientists Fund of the National Natural
Science Foundation of China (Grant No. 11204260), and the Program
for Changjiang Scholars and Innovative Research Team in University (IRT13093).\\
%+++++++++++++++++++++++++++++++++++++++++++++++++++++++++

\end{document}